\documentclass{emulateapj}
\newcommand{\til}{$\sim$}
\newcommand{\ergsqcmsec}{\thinspace\hbox{$\hbox{erg}\thinspace\hbox{cm}^{-2}
                \thinspace\hbox{s}^{-1}$}}

\def\spose#1{\hbox to 0pt{#1\hss}}
\def\simlt{\mathrel{\spose{\lower 3pt\hbox{$\mathchar"218$}}
     \raise 2.0pt\hbox{$\mathchar"13C$}}}
\def\simgt{\mathrel{\spose{\lower 3pt\hbox{$\mathchar"218$}}
     \raise 2.0pt\hbox{$\mathchar"13E$}}}

\newcommand{\ta}{SDSS\thinspace J1446+02}
\newcommand{\tb}{SDSS\thinspace J2050--05}
\newcommand{\tc}{SDSS\thinspace J2101+10}
\newcommand{\sat}{{\em XMM-Newton}}

\newcommand{\mek}{{\sc mekal}}
\def\today{\ifcase\month\or
January\or February\or March\or April\or May\or June\or
July\or August\or September\or October\or November\or December\fi
\space\number\day, \number\year}

\slugcomment{Accepted for publication in the Astronomical Journal 2006 August 29}

\shorttitle{Magnetic CV candidates from SDSS}
\shortauthors{Homer et al.}

\begin{document}


\title{Characterizing Three Candidate Magnetic CVs from SDSS: {\em XMM-Newton}
  and Optical Follow-up Observations$^{\star}$}


%\altaffiltext{$\dagger$}{Some of the results presented here were obtained with the MMT Observatory, a facility operated jointly by The
%University of Arizona and the Smithsonian Institution.}
\altaffiltext{$\star$}{Based on
observations obtained with the Sloan Digital Sky Survey and with the
 Apache Point
Observatory (APO) 3.5m telescope, which are owned and operated by the
Astrophysical Research Consortium (ARC)}
\author{Lee Homer\altaffilmark{1}, Paula Szkody\altaffilmark{1}, 
Arne Henden\altaffilmark{2,3,4}, Bing Chen\altaffilmark{5,6}, Gary D. Schmidt\altaffilmark{7},
Oliver J. Fraser\altaffilmark{1}, Andrew A. West\altaffilmark{1}}
% Gary Schmidt\altaffilmark{7}, Scott
%  F. Anderson\altaffilmark{1}, Nicole M. Silvestri\altaffilmark{1} and J. Brinkmann\altaffilmark{8}}
%\affil{Apache Point Observatory, 2001 Apache Point Road, P.O. Box 59, Sunspot, NM 88349-0059}
%\email{jb@apo.nmsu.edu}

\altaffiltext{1}{Department of Astronomy, University of Washington, Box 351580, Seattle, WA 98195, USA}
\email{homer@astro.washington.edu}
\altaffiltext{2}{Universities Space Research Association}
\altaffiltext{3}{US Naval Observatory, Flagstaff Station, P.O. Box 1149, Flagstaff, AZ 86002-1149, USA}
\altaffiltext{4}{American Association of Variable Star Observers, 25 Birch Street, Cambridge, MA 02138, USA}
\altaffiltext{5}{\sat\ Science Operations Centre, ESA/Vilspa, 28080, Madrid, Spain}
\altaffiltext{6}{VEGA IT GmbH, c/o European Space Operations Centre, Darmstadt, Germany}

\altaffiltext{7}{The University of Arizona, Steward Observatory, Tucson, AZ 85721, USA}
%\altaffiltext{8}{Apache Point Observatory, 2001 Apache Point Road, P.O. Box 59, Sunspot, NM 88349-0059, USA}
%\topmargin -0.5in
%\textheight 9.20in

%==============================================================================

\begin{abstract}
In the latest in our series of papers on \sat\ and ground-based optical
follow-up of new candidate magnetic cataclysmic variables (mCVs) found in the Sloan
Digital Sky Survey, we report classifications of three systems: SDSS
J144659.95+025330.3, SDSS J205017.84--053626.8, and  SDSS
J210131.26+105251.5.  Both the X-ray and optical fluxes of \ta\ are modulated on
a period of $48.7\pm0.5$ min, with the X-ray modulation showing the
characteristic energy dependence of  photo-electric absorption
seen in many intermediate polars (IP). 
% Indeed, our best fit model consists of two-temperature (80 eV and 58 keV)
% optically thin plasma emission (\mek\ model) together with a partial covering
% absorber.  A joint fit to spectra selected from the minimum and maximum flux
% intervals confirms that the reduction in X-ray flux is accounted for by the
% inclusion of a thick absorbing column of $4\times10^{22}$cm$^{-2}$, with a
% 61\% covering fraction.  
A longer period modulation and
radial velocity variation is also seen at around 4 hrs, though neither
dataset is long enough to constrain this longer, likely orbital, period
well. \tb\ appears to be an example of the most highly magnetized class of mCV, a
disk-less, stream-fed polar. Its 1.57 hr orbital period is well-constrained via
optical eclipse
timings; in the X-ray it shows both eclipses and an underlying strong, smooth modulation. In this case, broadly phase-resolved spectral fits indicate
  that this change in flux is the result of a varying normalization of the
  dominant component (a 41 keV \mek), plus the addition of a partial covering 
  absorber during the
  lower flux interval. 
%These data indicate that only one pole accretion is occurring, that this pole
%is not self-eclipsed by the white dwarf, and that aspect changes dominate the
%variations in observed flux. In addition, the requirement for a partial
%covering absorber indicates that at limited phases, the accretion stream
%intersects the line-of-sight; highly likely for such a high-inclination system.
  \tc\ is a more perplexing system to categorize: its
  X-ray and optical fluxes exhibit no large periodic modulations; there are only
  barely detectable changes in the velocity structure of its optical emission
  lines; the X-ray 
  spectra require only absorption by the
  interstellar medium; and the temperatures of the \mek\ fits are low, with
  maximum temperature components of either 10 or 25 keV.  We conclude that \tc\ can
  not be an IP, nor likely a polar, but is rather most likely a disc accretor-- a low
  inclination SW Sex star. 
% The only other class of CV to exhibit such strong, high excitation lines like
%  \ion{He}{2} are the SW Sex stars.  
%We suggest that it may well be an SW Sex star, but
%  viewed at a much lower inclination than the rest of the class, thus explaining
%  the barely detectable changes in the velocity structure of its emission lines,
%  and lack of clear orbital modulation.
% What variations we do see in our X-ray and optical lightcurves appear on
% timescales around 100 min, though in all cases this is comparable to the data span.  \tc\ in particular will benefit from much more extensive ground-based observation.

\end{abstract}

\section{Introduction}

The Sloan Digital Sky Survey \citep[SDSS;][]{york00} has provided a wealth
of new cataclysmic variables; the close binary systems with active accretion
from a late main sequence star to a white dwarf \citep[reviewed in][]{warn95}.
Due to its sensitivity down to $\sim$21st mag, SDSS is especially suited to the 
discovery of faint CVs with low accretion rates and short orbital periods.
Included in the discoveries are dozens of systems with a noticeable emission
line of \ion{He}{2}, which is often a signature of a white dwarf with a 
high magnetic field.
The highest field systems (polars) have no 
accretion disk and the white dwarf spin is synchronized to the orbit.
The intermediate polars (IPs) usually have some outer disk, with magnetic curtains
channelling the inner disk material to the magnetic poles of the white dwarf
which is spinning faster than the orbital timescale. Polars can be
identified by circular polarization, cyclotron harmonics and/or
Zeeman splitting \citep[review in][]{wick00}, while
IPs are found through the detection of their spin and orbital periods. 
In addition, there is a class of high accretion rate disk systems with possible
low magnetic field white dwarfs \citep{rodr01}, termed
SW Sex stars, which can also show strong \ion{He}{2} lines in their optical
spectra.  

For the last few years, we have used \sat\ to identify the nature
of the CVs with \ion{He}{2} and to explore the nature of X-ray heating
in these systems. Polars generally exhibit hard X-rays from the
accretion shock and soft X-rays from the white dwarf surface heated by
the hard X-rays. The ratio of soft to hard X-rays shows a dramatic
change with magnetic field strength and accretion rate. The IPs are
generally hard X-ray emitters with large absorption effects from the
accretion curtains. Usually the spin period is much more pronounced in
the X-ray than the optical.  The SW Sex stars usually show
no X-ray variation and are hard X-ray emitters with low absorption.

The 3 sources in this study are all about 18th mag in the SDSS $g$ filter and
were all identified as potential magnetic CVs
in \citet{szko03}. SDSS\thinspace J144659.95+025330.3 (hereafter \ta) 
showed no polarization and limited time-resolved spectra could not
determine the orbital period. SDSS\thinspace J205017.85--053626.8 (\tb)
has the strongest \ion{He}{2} line of the 3 objects, with a
peak flux comparable to H$\beta$ and showed evidence
of high and low states of mass transfer. 
A short span of time-resolved spectroscopy
showed an orbital period near 2 hrs. Followup high speed photometry by
\citet{woud04} revealed brief (260s), deep (1.5 mag) eclipses and  determined a
precise period of 1.57 hrs. 
The third system, SDSS\thinspace J210131.26+105251.5 (\tc) has very strong
Balmer lines but did not show any velocity variation during 1.3 hr
of time-resolved spectroscopy.

The {\em XMM-Newton} data obtained on these 3 sources, combined with
additional photometry, spectroscopy and spectropolarimetry has
allowed us to determine that \ta\ is an IP, \tb\ is a likely
polar, while \tc\ escapes an easy classification, but appears most likely to be an SW Sex star.
Our observations
and results are described below.

\section{Observations}
\begin{deluxetable*}{lllcll}
%\begin{deluxetable}{lllcll}
%\tablenum{1}
%\rotate
\tablewidth{0pt}
\tabletypesize\small
\tablecaption{Observation Summary\label{tab:obslog}
}
\tablehead{\colhead{SDSS J} &\colhead{UT Date} & \colhead{Obs} & \colhead{UT Time} & \colhead{Characteristics\tablenotemark{a}} &
\colhead{Comments} }
\startdata
\ta & 2002 May 11 & SO: Bok (2.3m) & \nodata & $v=+0.38$\%& 3ks exposure, spectropolarimetry\\
&  2003 Apr 27 & APO & 05:43 -- 09:47&\nodata  & $23\times 600$s spectra\\
 & 2003 May 07 & NOFS & 06:39 -- 11:34 &\til$V=17.9$ & open filter photometry\\
&2004 Jan 30& \sat: &&&\\
&& EPIC-pn & 04:17 -- 05:32 &0.2 cts s$^{-1}$&only 3.0ks live time\tablenotemark{b,c} \\
&& EPIC-MOS1/2 & 03:55 -- 05:18 & 0.1 cts s$^{-1}$&only 4.5ks live time\tablenotemark{c}\\
&& OM & 04:03 -- 08:10 & $B=18.3$ &13.9ks duration\\
& & NOFS & 09:13 -- 13:18 &\til$V=18.2$ & open filter photometry\\
&2005 Jan 12& \sat: &&&\\
&& EPIC-pn &06:22 -- 08:09  & 0.20 cts s$^{-1}$&5.6ks live time\\
&& EPIC-MOS1/2 &06:00 -- 08:14  & 0.10 cts s$^{-1}$&7.7ks live time\\
&& OM & 06:08 -- 08:16 & $B=18.2$ &5.6ks duration\\

\tb & 2003 May 29  & MMT & \nodata  & $v=+1.06$\% & 1.2ks exposure, spectropolarimetry\\
& 2003 May 30 & MMT & \nodata  &$v=-0.23$\%& 2.4ks exposure, spectropolarimetry \\
& 2004 May 14 & SO: Bok (2.3m) & \nodata  & $v=-0.31$\%& 1.2ks exposure, spectropolarimetry \\
&2004 Oct 18& \sat: &&&\\ 
&& EPIC-pn &10:13 -- 13:17  & 0.67 cts s$^{-1}$&8.7ks live time\tablenotemark{b} \\
&& EPIC-MOS1/2 &09:50 -- 13:22  & 0.21  cts s$^{-1}$&12ks live time \\
&& OM & 06:08 -- 08:16 & $B=18.4$ &8.4ks duration\\
& 2004 Oct 21 & NOFS & 01:33 -- 04:34 &\til$V=18.0$ & open filter photometry\\

\tc & 2003 May 29 & SO & \nodata & $v=+0.03$\%& 1.6ks exposure, spectropolarimetry\\
& 2003 Sep 22 & SO: Bok (2.3m) & \nodata & $v=+0.10$\%& 1.6ks exp., $R$-band imaging polarimetry\\
& 2004 May 14 & APO & 10:47 -- 11:04 &\nodata & 1ks spectrum\\
 & 2004 May 19  &  NOFS & 08:01 -- 11:35  &\til$V=18.9$ & open filter photometry\\
&2004 May 19& \sat: &&&\\ 
&& EPIC-pn &16:57 -- 18:13  &0.64 cts s$^{-1}$&4.1ks live time\tablenotemark{b} \\
&& EPIC-MOS1/2 &16:35 -- 18:18  & 0.23 cts s$^{-1}$&6.1ks live time \\
&& OM & 16:43 -- 18:20 & $B=19.0$ &5.5ks duration\\

\enddata
\tablenotetext{a}{The open filter photometry from NOFS has an estimated
  equivalent $V$ zero-point, while for spectra the flux density at \til5500\AA\ is used. The \sat\ count rates are average values for each
  observation for a single detector.}
\tablenotetext{b}{The live time of the X-ray CCD detectors refers to the sum of the good-time intervals, less any dead time.  It
  is typically much less than the difference of observation start and stop times.}
\tablenotetext{c}{These X-ray exposures were curtailed by severe particle background flaring.}
\end{deluxetable*}
%\end{deluxetable}

\subsection{X-ray}
Table~\ref{tab:obslog} summarizes our X-ray and optical observations. \ta\
 was observed twice by \sat, as the 2004 January 30
observation was badly affected by high charged particle background throughout, 
and  the X-ray exposures were curtailed after only a few
kiloseconds. Since the X-ray instrumentation \citep[comprising three X-ray telescopes, backed by the two MOS and one pn EPIC CCD
cameras,][]{turn01} was set as the prime instrument, an automatic reobservation was 
triggered.  However, the 2004 optical data from \sat's optical monitor \citep[OM,
][]{stru01} were good throughout the scheduled \til13 ks observation.  In the 
second observation on 2005 January 12, good data were obtained
from all three EPIC spectro-imagers, but this time the OM only succeeded for 2 
out of an expected 4 exposures. We note that for both
observations no useful data were available from the two Reflection Grating 
Spectrographs \citep{denH01} arrayed in the optical path of the MOS
detectors, as the source is too faint to furnish adequate signal in the dispersed
spectra.  The initial visits made to \tb\ and \tc\ were highly successful, with 
only a little background flaring in the last 3 ks of the 12 ks
observation of \tb, and none affecting that of \tc. Useful data were therefore 
obtained from EPIC imaging and the OM, though once again neither
target was bright enough for the RGS.

% In summary, in the analysis presented below we used all OM lightcurve data from both \sat\ observations, but only X-ray data from the EPIC cameras from the 2005 return visit. 

For the extraction of spectra and lightcurves from the \sat\ data we used the 
tools available in the  Science
Analysis System (SAS\setcounter{footnote}{8}
\footnote{Available from
http://xmm.vilspa.esa.es/external/ xmm\_sw\_cal/sas.shtml}) version 6.5.0, with  calibration files current to 2005
December 15, and followed the standard protocol as given by the ESA \sat\ 
website and the ABC
guide\footnote{http://heasarc.gsfc.nasa.gov/docs/xmm/abc/abc.html} from the US GOF. In all cases we produced new event list files from the
Observation Data Files to incorporate the latest calibration updates and then
filtered them with the standard canned
expressions. We defined circular source extraction regions (radii of 480 and 
360 pixels for the MOS and pn camera respectively), centered on the centroid 
of the source, enclosing  $\sim80\%$ of the energy to
optimize for signal-to-noise. Simple annular background extraction regions were
defined on the same central MOS chip for those 2 cameras; for
the pn we used  adjacent rectangular regions at similar detector Y
  locations to the target. We also reprocessed the OM data with  {\tt omfchain}
 extracting lightcurves using more appropriate (for our
  relatively faint targets) source aperture and background regions to maximize 
signal-to-noise.  We also chose binning to match
 that of the ground-based photometry, and lastly, to aid comparison, converted 
count rates to $B$ magnitudes.

The SAS task {\tt evselect} performed the extractions of both X-ray spectra and lightcurves for source and background.  For \ta\ and \tb\ further time filtering was
applied to the event lists prior to spectral extraction to excise intervals of 
high X-ray background; the revised good time intervals (GTIs) were
generated by setting limits on the count rate in hard ($>10$keV) lightcurves for the entire detector.  We also restricted acceptable events to
singles in the pn (pattern=0) but up to quadruples in the MOS (pattern$\leq12$) to further improve energy resolution. In contrast, in extracting
lightcurves only the standard GTIs were invoked, and we kept both singles and doubles for the pn (pattern$\leq4$), to both maximize lightcurve
coverage and signal-to-noise. For \tc\ we also restricted the energy range to below 2.5 keV, where the source
contributions dominate.  SAS tasks {\tt rmfgen}, {\tt arfgen} and {\tt backscale} generated appropriate
redistribution matrix (rmf) and ancillary response files (arf) and calculated the relative scaling of source to background. 

The final steps in data reduction utilized general purpose utilities in the FTOOLS\footnote{http://heasarc.gsfc.nasa.gov/lheasoft/ftools/}
software suite: {\tt grppha} grouped the spectral bins and associated various
files ready for spectral analysis in XSPEC; {\tt lcmath} created
background subtracted lightcurves, and combined the two MOS lightcurves, while {\tt earth2sun} applied a correction to the time stamps for the
solar system barycenter.

\subsection{Optical}

Ground based photometry was obtained for all three sources using the US
Naval Observatory Flagstaff Station (NOFS) 1m telescope and a 2048 $\times$2048
CCD. An open filter was used which is close to a $V$ response and the magnitudes 
were calibrated from separate nights of all-sky photometry  with Landolt
standards. Light curves were made by using differential photometry with respect
to comparison stars in each field and the magnitudes measured using
IRAF\footnote{IRAF (Image Reduction and Analysis Facility) is distributed by 
the National Optical Astronomy Observatories, which are operated
  by the Association of Universities for Research in Astronomy (AURA) Inc., 
under cooperative agreement with the National Science Foundation} routines.
For \ta, 4 hrs of photometry began about one hour after the \sat\ observation
ended on 2004 Jan 30. No data were obtained for the second observation but
an additional 5 hrs of photometry from 2003 May 07 was used to establish the 
optical period.
For \tb, 3 hrs of photometry were obtained 3 days after the \sat\ observations,
while the 3.5 hrs of observations for \tc\ took place 5.5 hrs prior to the
start of \sat\ coverage. In all cases, the ground-based measurements agreed with the
OM in showing the systems were all in their normal state of accretion.

In order to determine the orbital period for \ta, 4 hrs of time-resolved 
spectra were obtained  on 2003 Apr 27. The Double Imaging Spectrograph
(DIS) was used on the 3.5m telescope of the Apache Point Observatory (APO).
Twenty three 10 min blue and red spectra were obtained with a 1.5\arcsec\ slit,
covering the regions 4200-5100\AA\ and 6300-7200\AA\ with a resolution of
about 2\AA. IRAF routines were used to calibrate the spectra for wavelength
and flux using standards from the night. Velocities were measured for the prominent emission 
lines using the centroid (``e'') routine in the IRAF splot package;  a double-Gaussian method \citep{shaf83}
was tried as well. A single spectrum of \tc\ was also obtained at APO 5 nights before the \sat\
observation. The Balmer and \ion{He}{2} strengths are very similar to the
SDSS spectrum shown in \citet{szko03}.
 
Circular spectropolarimetry was also performed for the three systems
in a search for evidence of magnetic fields.  The CCD
spectropolarimeter SPOL was used \citep{schm92} on the Steward
Observatory 2.3~m Bok telescope and the 6.5~m MMT, as indicated in
Table 1.  All measurements on SDSS J1446+02 and SDSS J2101+10, and
two of the three on SDSS J2050$-$05, yielded spectrum-averaged values
$|v| = |V/I| < 0.4\%$.  Each result is well within $3\sigma$ of zero,
and thus consistent with being unpolarized.  However, a third epoch
on SDSS J2050$-$05, obtained on 2003 May 29 in good observing
conditions at the MMT, yielded $|v|=+1.06\%$.  The S/N of these data
is sufficient to reveal that the circular polarization rises
continuously from $v\sim0$ at $\sim$4200~\AA\ to $v\sim3\%$ beyond
$\lambda=8000$~\AA.  This and other evidence for polarization by
cyclotron emission in SDSS J2050$-$05 is discussed in $\S$4.2.

\section{Analysis and results}
\subsection{X-ray Spectral Fitting}
The extracted spectra were binned at $>$20 counts/bin, to facilitate the use of
$\chi^2$ statistics to find the best model fits.  For all sources the background
contributed less than 3\% of the flux within the source aperture, hence simpler fitting of
background subtracted spectra was deemed acceptable. At the very lowest energies the calibration of the EPIC detectors remains uncertain; following the
latest guidelines we restricted our fitting to $>0.2$keV for the MOS and
$>0.15$keV for the pn. In every case, we performed  joint fits to the data from all
three X-ray cameras simultaneously, where all model parameters were fixed apart from the
relative normalization of the pn relative to the MOS (which were assumed to be
identical).  In the cases where we separated the data into two phase or time
bins, we then had 6 datasets.  Starting from the case where all relative
normalizations, as well as other model parameters, were kept constant, we then
allowed normalizations to vary (but keeping the MOS to pn ratios constant), and
then other model parameters like temperature, $F$-testing to see whether the
additional degrees of freedom were statistically warranted each time.  The best
fits we present in figures and tables are those where the model has the fewest
free parameters (i.e. leaves the most
degrees of freedom); any further freeing of parameters did not then signficantly
reduce the (reduced) $\chi^2$ values.

Furthermore, in finding the best fit models we always started from the simplest single emission component, solely absorbed by interstellar dust, with an initial
value for $N_H$ as predicted by dust maps for the appropriate line of sight through the Galaxy. Typically, we used a single bremsstrahlung or a
version including explicit line emission-- the XSPEC model \mek -- as expected for the optically thin thermally-emitting plasma
encountered in CVs. In all cases, these simplest models failed to provide a good fit, hence we moved onto a variety of more complex
combinations, e.g. multi-temperature or two-temperature thermal plasmas, additional soft blackbody components (as emitted by the heated polar caps
in polars) and the effects of local partial
covering absorption (i.e. due to obscuration by the accretion stream and/or curtain). We discuss the details of the best fits, as we report the
results of our various analyses for each target in turn. 

\subsection{\ta}
%\begin{deluxetable}{llllll}
\begin{deluxetable*}{llllll}
%\tablenum{1}
\tablecolumns{6} 
\tablewidth{0pt}
\tablecaption{X-ray Spectral Fits for \ta\label{tab:1446xfits}}
\tablehead{ \colhead{Model} &\colhead{reduced} & \colhead{$N_H$} & \multicolumn{2}{c}{Partial covering} &\colhead{Emission} \\
  &\colhead{$\chi^2$}   &\colhead{$\times10^{20}$}& \colhead{$N_H$ } & \colhead{frac.}&  \colhead{model parameters}\\
 &\colhead{(d.o.f.)}&cm$^{-2}$&\colhead{($\times10^{22}$cm$^{-2}$)}&
}
\startdata
bremss &1.5 (139)&$6.4\tablenotemark{a}$&\nodata&\nodata&$kT=200$ (pegged)\\
power law&1.11 (136)&$1.4^{+0.8}_{-1.0}$ &$0.8\pm0.2$&$0.41^{+0.04}_{-0.07}$&$\Gamma=1.2\pm0.2$\\
multi-$T$ \mek\tablenotemark{b}&1.14 (137)&$1.6\pm0.7$&$1.3\pm0.2$&$0.64\pm0.02$&$kT_{\rm max}=60$ keV (fixed\tablenotemark{c})\\
&&&&&$\alpha=1.0$ (fixed)\\ 
2 $T$
\mek&1.10(134)&$4^{+27}_{-2}$&$0.9^{+3.9}_{-0.2}$&$0.47^{+0.05}_{-0.04}$&$kT_1=130^{+2}_{-4}\;$eV\\
&&&&&$kT_2=80^{+0}_{-19}$ keV\\
\cutinhead{{\em Joint fit to phase-selected ``maximum'' and ``minimum'' flux intervals:}}
2 $T$ \mek& 1.08 (104)& $13^{+2}_{-3}$&none (max)& none (max)
&$kT_1=81^{+4}_{-0}$ eV\\
&&& $4.4^{+1.6}_{-0.8}$(min)& $0.61\pm0.03$(min)&$kT_2=58^{+22}_{-19}$ keV\\
\enddata
\tablenotetext{a}{This fit did not converge well, hence no error estimates are available.}
\tablenotetext{b}{We used the XSPEC model {\sc cemekl}, in which a fine grid of \mek\ models are coadded, with the emission
measures following a power-law in temperature with
  index $\alpha$ and up to $kT_{\rm max}$ (i.e. normalizations scale as $(T/T{\rm max})^\alpha)$.}
\tablenotetext{c}{$kT_{\rm max}$ fixed to plausible value, free fit
does not converge.}
%\tablenotetext{c}{Model parameters without quoted uncertainties have been fixed at the value given.}
\end{deluxetable*}
%\end{deluxetable}

\begin{figure}[!tb]
%\resizebox{.49\textwidth}{!}{\rotatebox{0}{\plotone{1446_RV.eps}}}
\resizebox{.49\textwidth}{!}{\rotatebox{0}{\plotone{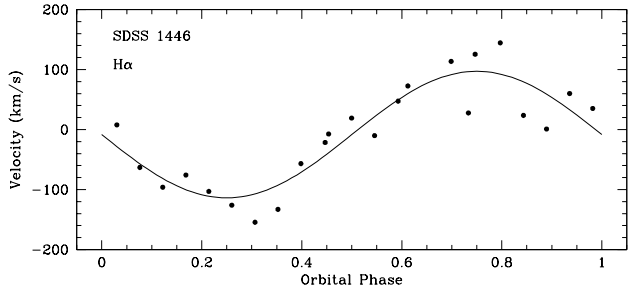}}}
\caption{Radial velocities for the H$\alpha$ emission line in \ta, phase folded
  on the best fit period, $P=3.8\pm0.3$ hrs.  Overplotted is
  the sinusoid fit with $K= 105\pm11$ km s$^{-1}$ and $\gamma=-8\pm5$ km s$^{-1}$. \label{fig:1446RV}}
\end{figure}

\subsubsection{Search for periodicities}
From our APO spectroscopy run on \ta, we generated radial velocity curves for
the prominent emission lines.  A least square sine fit to the velocities was then used to
determine the systemic velocity, the semi-amplitude, the orbital period and
the phase (based on the red-to-blue crossing). While the H$\alpha$, H$\beta$,
H$\gamma$ and \ion{He}{2} lines were all measured, there were large deviations
with respect to the best fit sine wave in all cases (total sigma of the fit of 34, 46, 75 and 73 km/s respectively), hence we
only consider the fit to the H$\alpha$ curve in any detail (shown in fig.~\ref{fig:1446RV}). All lines gave a
period solution near 4 hrs, which is very close to the length of the dataset.
The H$\alpha$ solution shown in figure~\ref{fig:1446RV} is for a period of $3.8\pm0.3$ hrs,
$K= 105\pm11$ km s$^{-1}$ and $\gamma$ of $-8\pm5$ km s$^{-1}$, with red-to-blue crossing (phase 0) at 7:30 UT on 
2003 Apr 27. Because the period is so close to
the length of the observation, it is not a robust determination, but useful
in showing that the orbital period is not short (i.e. under 2 hrs).

\begin{figure}[!tb]
\resizebox{.49\textwidth}{!}{\rotatebox{-90}{\plotone{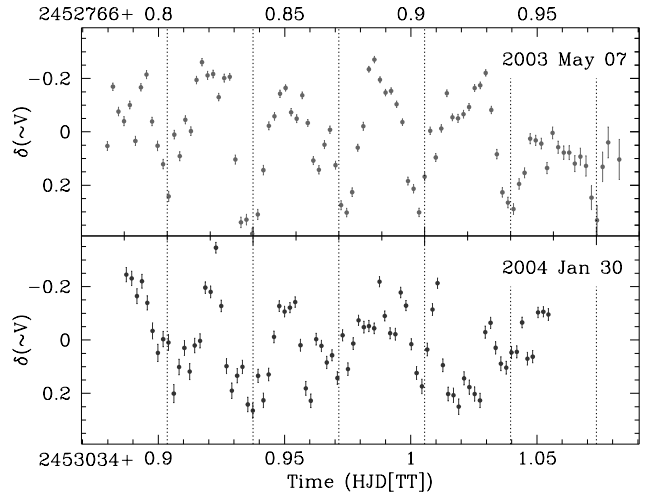}}}
%\resizebox{.49\textwidth}{!}{\rotatebox{-90}{\plotone{1446_optlcs.eps}}}
\caption{The NOFS \til$V$ lightcurves from the two different observations of
  \ta\ are plotted using a common time-scaling.  We also over-plot vertical bars
to mark the length of each spin cycle; comparison of the variations during each successive spin cycle shows how much more regular they are in 2003 May
07  than in 2004 Jan 30; possibly enhanced flickering is responsible for the increased scatter in the latter curve. \label{fig:1446optlcs}}
\end{figure}

\begin{figure}[!htb]
%\resizebox{.47\textwidth}{!}{\rotatebox{0}{\plotone{1446_bfs.eps}}}
\resizebox{.47\textwidth}{!}{\rotatebox{0}{\plotone{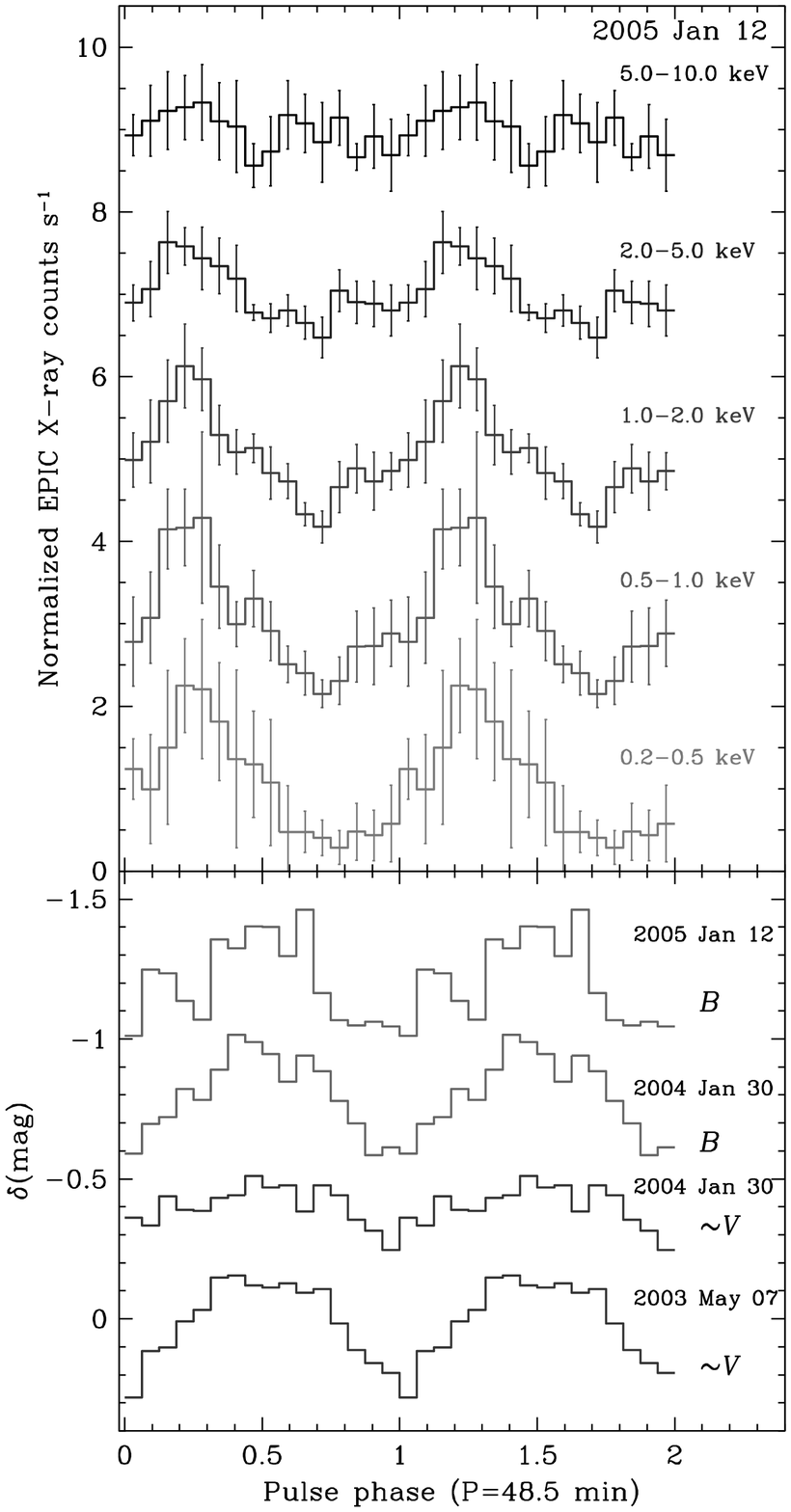}}}
\caption{White Dwarf spin phase folded, and binned light curves for \ta. {\em Upper:} the energy-selected EPIC X-ray light curves; note the
marked change in amplitude, increasing with decreasing energy. {\em Lower:}
the optical light curves from the two \sat/OM ( in $B$) and two NOFS observations (in an approximate $V$ band); in the NOFS light curves the change in
amplitude, due to the variation in scatter about the mean curve is made very
apparent. Note: (i) the simultaneous X-ray and OM 2005 Jan 12 lightcurves share
a common ephemeris, arbitarily chosen to place optical minimum at phase=0.0,
the phasing of each of the other optical lightcurves has simply been chosen to align
their optical minima; (ii) for clarity each of these lightcurves has been offset
vertically, with 2 normalized X-ray counts and 0.4 mags respectively. \label{fig:1446bfs}}
\end{figure}

For \ta\ we also possess four optical lightcurves (two from the ground and two
from the OM), plus those from the EPIC X-ray cameras.  To maximize signal-to-noise we summed the results from
the two MOS; however, since these are always of longer duration than the pn, we
did not combine pn and MOS at this stage. We performed a Lomb-Scargle
periodogram \citep{scar82} analysis of each of our lightcurves.  In every case, a significant peak was found at a period of 49 min, which we identify as the
spin period of the white dwarf.  In addition, longer term variations were apparent in the optical lightcurves with peaks corresponding to around
4 hrs, close to the signal found in the time-resolved APO spectra. To determine the periods most accurately, we used sinusoidal fitting to the longest 2003 May 7 NOFS
lightcurve, with a sinusoid plus first harmonic model for the pulse, and a simple sinusoid for the longer period variation.  This yielded a
period of $0.0338\pm0.004$d ($48.7\pm0.5$ min) for the pulse, and  $0.165\pm0.015$d ($4.0\pm0.4$hr) for the longer period, which we presume to
be the orbital period of the system. Once again we caution that the 2003 May curve  provides only 1.3 cycles coverage of our tentative orbital period,
and it remains poorly constrained. Nevertheless, we then fit the other three optical curves
with  sinusoids constrained to the pulse and orbital periods, and subtracted the
latter. In fig.~\ref{fig:1446optlcs} we show the two resulting NOFS
lightcurves; the pulse profile is obvious in most cycles in the 2003 May curve,
whereas in 2004 Jan there is far more scatter, especially away from minimum
light.  This may simply be due to varying amounts of flickering, perhaps
indicative of small changes in accretion state. Lastly, we phase folded (and
binned) the light curves on the pulse period
to examine its profile in greater detail. The binned results are shown (along with the X-ray) in  Fig.~\ref{fig:1446bfs}. The profile is
clearly far from sinusoidal, having a relatively broad and flat-topped peak, and narrower V-shaped minimum.

Before phase-folding the X-ray data, we created five different energy selected lightcurves, in order to investigate any dependence of the pulse
profile on energy. The final binned lightcurves shown in Fig.~\ref{fig:1446bfs} were constructed by summing the {\em folded} and binned results for
all three X-ray cameras to maximize signal-to-noise.  Perhaps not too
surprising, the profile of the X-ray pulse is very different from the optical,
but it is noteworthy that the X-ray and optical are neither in phase nor
anti-phased.  Instead, the peak of the X-ray roughly leads that of the optical by \til0.3 in phase.  The X-ray pulse
is also energy dependent, with the largest amplitude (and symmetrical peaked pulse) in the two lowest energy bins, the amplitude then decreases
with energy, until above 5 keV
there is no significant variation. 
% We also note the possible appearance of a secondary peak leading the main peak
% by \til0.2 in phase, though below 1 keV it is little more than an asymmetric shoulder in the main peak's profile.

\begin{figure*}[!tb]
%\resizebox{.77\textwidth}{!}{\rotatebox{-90}{\plotone{1446_xspecfit.eps}}}
\resizebox{.77\textwidth}{!}{\rotatebox{-90}{\plotone{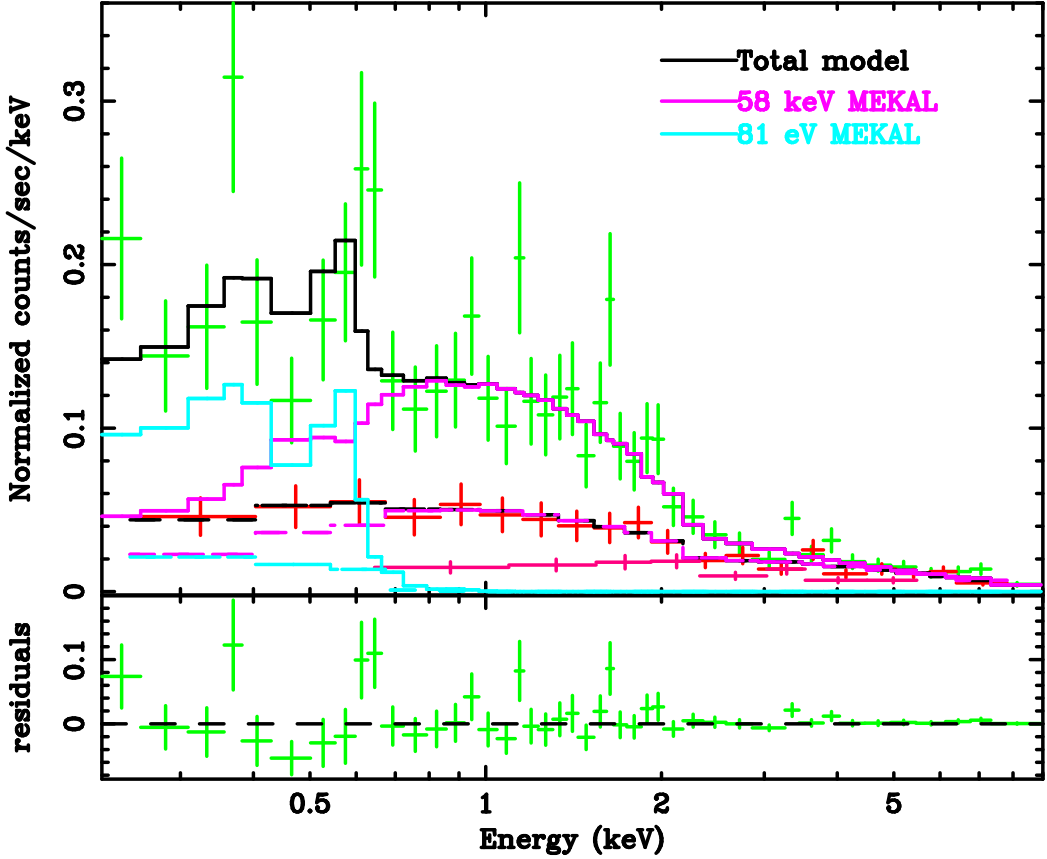}}}
\caption{{\em Upper panel:} \sat/EPIC-pn spectra for \ta\ from the ``max'' (green or black points) and ``min''
  (red or dark grey) spin phase intervals; the MOS data, used in the full joint
  fits, have been omitted for clarity. The best fit two temperature \mek\  model is
  over-plotted for each (solid and dashed respectively); we also plot the two
  component contributions. {\em Lower panel:} the residuals to the fit of the total model.\label{fig:1446xspec}}
\end{figure*}

\subsubsection{Spectral variations with phase}
 The variation in pulse amplitude with energy is exactly that which we
would expect if obscuration by a local absorber is responsible.  In fitting the X-ray spectra, we first found a fit to the entire dataset, but
then applied this model separately to spectra phase-selected from the maximum
 ($\phi=0.05-0.55$) and minimum ($\phi=0.55-1.05$) flux intervals of the pulse profile. The details of the
fits are given in Table~\ref{tab:1446xfits}.  The model that has the best fit 
consists of a two-temperature thermal plasma (\mek), absorbed both by a small Galactic column, but also by a much larger variable partial
covering absorber. The fit to the combined spectrum achieved a barely acceptable reduced  $\chi^2=1.1$ for 134 degrees of freedom (d.o.f.),
though it does better than any other model.  This model has a well-constrained cool plasma of $kT=130^{+2}_{-4}$eV contributing unresolved soft emission line structure plus a very hot component
whose temperature pegged at 80 keV, the upper limit for the \mek\ model.
Furthermore, the values of the Galactic and local columns
were very poorly constrained. This is probably an effect of trying to fit  a highly
phase-variable spectrum with a single model (even this fairly complex one). Indeed, after separating the data into ``max'' and ``min'' subsets, a joint fit to these spectra was a significant
improvement (see fig.~\ref{fig:1446xspec}). There was not a significant change in the reduced $\chi^2$, but we were now able to find  well-constrained values for the absorbing
columns and a physically plausible temperature for
the hottest \mek\ component of $kT=58^{+22}_{-19}$keV. Once again the fit
required a very cool \mek\ to account for
line structure
in the 0.3--0.4 keV range, plus \ion{O}{7} emission at 0.56 keV; indeed this
parameter now pegged at its minimum value of 81 eV, indicating that some
problems still remain (but the need for a very cool plasma is less problematic
than an unphysically hot one, given the physical conditions in the expected
complex multi-temperature accretion column).  In this fit, the partial covering absorber parameters for the ``max'' tended to zero, hence we fixed them as such; the ``min''
interval as expected requires a thick column around $4\times10^{22}$cm$^{-2}$, with 61\%
covering fraction. In our joint fit we found no statistical support for varying any
of the normalization parameters; the change between ``max'' and ``min'' is
entirely accounted for by the introduction of the thick partial covering
absorber. Therefore, the fully unabsorbed flux (0.01--10 keV) remains constant
at $3\times10^{-11}\ergsqcmsec$, whereas the affect of the absorption leads to clear energy-dependent changes in the observed fluxes: a drop of 60\% ($4.6-1.8\times10^{-13}\ergsqcmsec$) in the
softest  0.2--2.0 keV band; a decrease of only 30\%
($7.6-5.6\times10^{-13}\ergsqcmsec$) in the 2.0--5.0 keV; and no significant
change in the flux above 5 keV.

\begin{figure*}[!tb]
%\resizebox{.77\textwidth}{!}{\rotatebox{0}{\plotone{2050_fds.eps}}}
\resizebox{.77\textwidth}{!}{\rotatebox{0}{\plotone{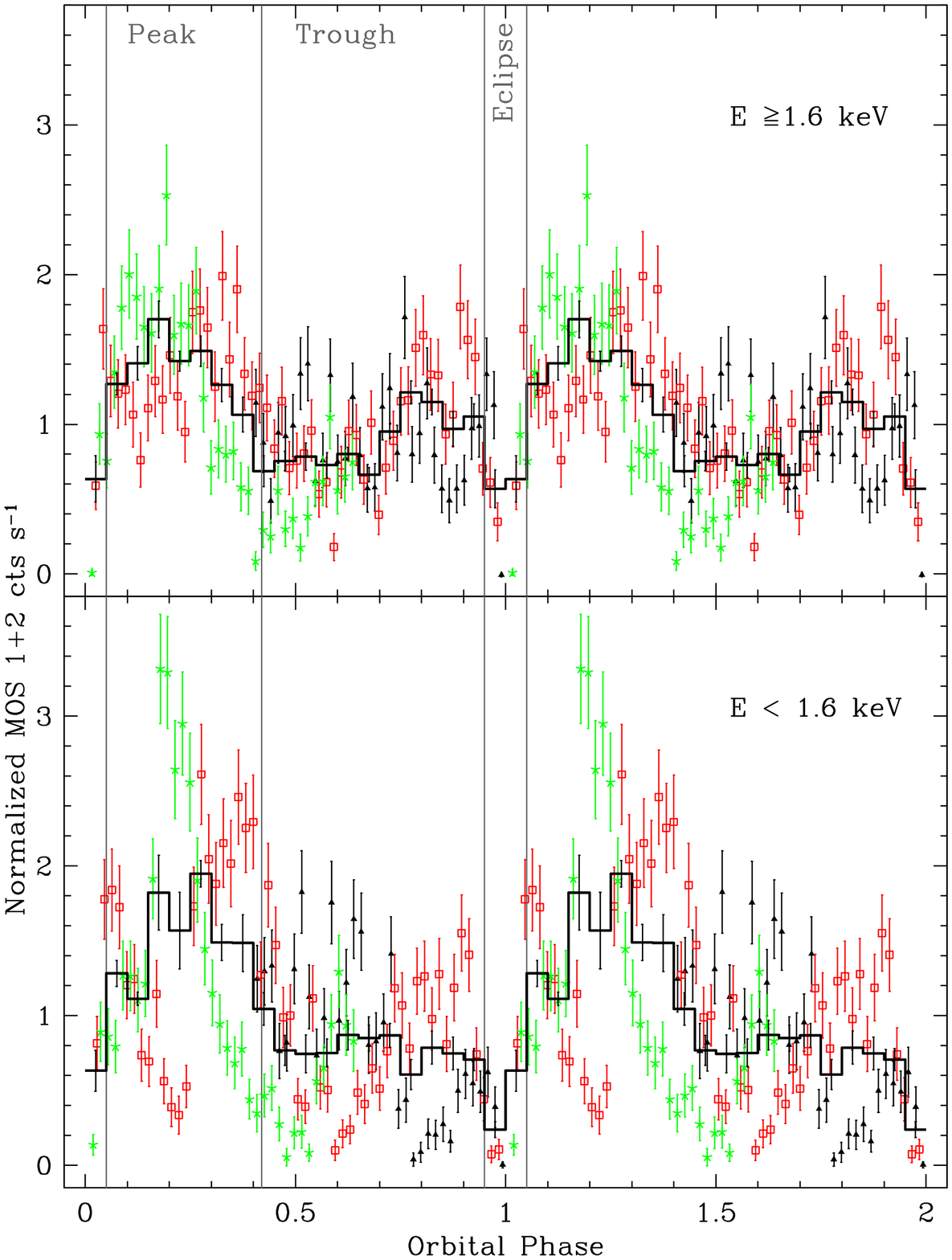}}}
\caption{Energy-selected \sat/EPIC-MOS (combined) X-ray lightcurves of \tb, phase-folded on our revised ephemeris: {\em (top panel)} a hard band with
  $E\ge1.6$keV; {\em (bottom panel)} soft band with $E<1.6$ keV. Each of the
  three cycles covered is shown with a different symbol and grey shade (color
  in electronic edition) in order to show how much of the variability is apparently random flickering, and is more pronounced in the soft
  energy band. The average orbital modulation is brought out by binning the
  data; this curve is over-plotted in solid black. The vertical grey bars and
  labelling indicate the three phase intervals into which the data were
  subdivided to study variations in the spectra.\label{fig:2050fds}}
\end{figure*}

\newpage
\subsection{\tb}
The eclipses seen in the optical photometry of \citet{woud04} are also very apparent in the X-ray lightcurves, and in our $B$-band OM results. Their ephemeris is sufficiently precise that there is merely a 0.01 cycle uncertainty at the time of our \sat\ observations. Adding our
own eclipse center measurements, we are able to even further refine the ephemeris to: \[{\rm
  HJD_{min}(TT)}=2453296.29816(6)+0.06542463(1)\ast E \;\;\;{\rm d}\] where the numbers in parenthesis indicate the $1\sigma$ uncertainty in the
last digit.  Also, note that the time is given in Terrestrial time
(UT=TT+64.184~s at the present epoch).  

\begin{figure*}[!tb]
%\resizebox{.77\textwidth}{!}{\rotatebox{-90}{\plotone{2050_xspecfit.eps}}}
\resizebox{.77\textwidth}{!}{\rotatebox{-90}{\plotone{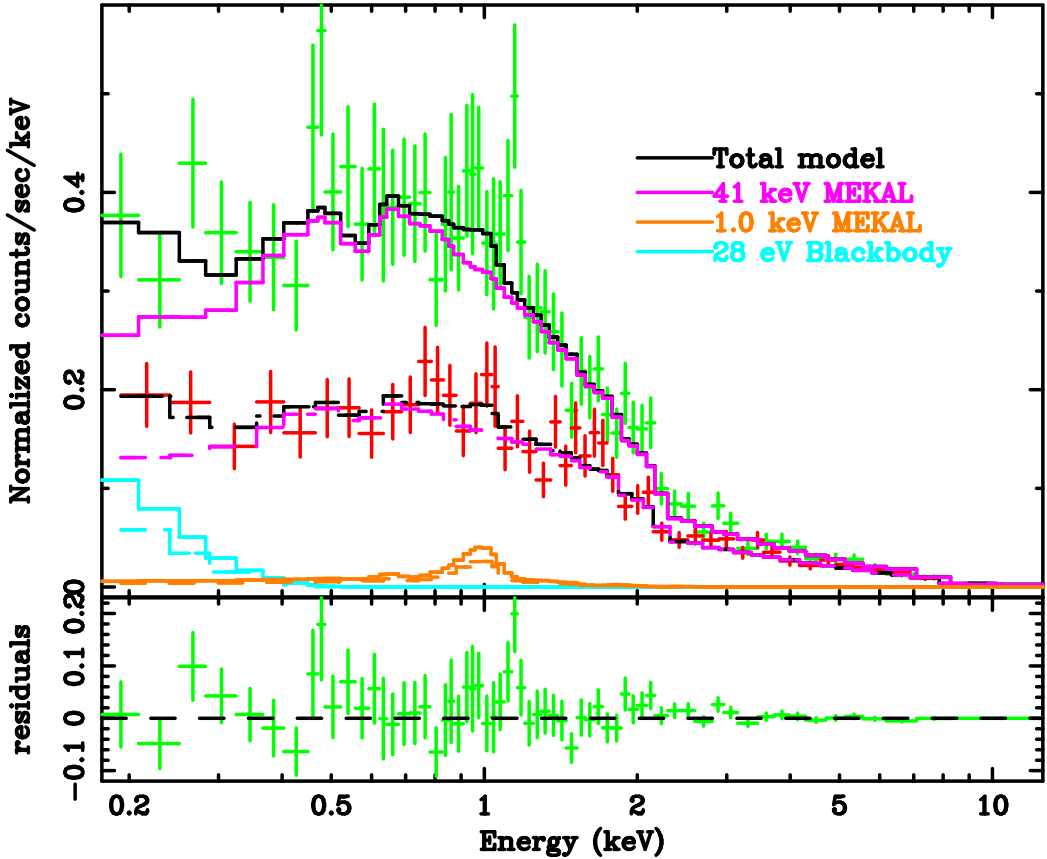}}}
\caption{{\em Upper panel:}\sat/EPIC-pn spectra for \tb\ from the ``peak'' (black or green points)
  and ``trough'' (red or dark grey) orbital phase intervals; the MOS data, used in the full joint
  fits, have been omitted for clarity. The best fit two temperature \mek\ plus
  blackbody model is over-plotted for each (solid and dashed respectively); we also plot the various component contributions. The hotter
  emitting gas provides the dominant contribution, and the change in overall
  flux between the intervals is due to a combination of its changing
  normalization and the extra absorption from a dense partial covering
  component. {\em Lower panel:} the residuals to the fit of the total model.\label{fig:2050xspec}}
\end{figure*}

Besides the eclipses, the X-ray lightcurves exhibit large amplitude but apparently irregular flaring behaviour.  We divided the X-ray data into
two energy bins, above and below 1.6 keV, roughly encompassing the same count rates, then phase-folded the lightcurves and we plot them in fig.~\ref{fig:2050fds}, where
we also plot each cycle of data with a different symbol (and color in the
electronic edition).  Although this energy division is not physically motivated,
it does roughly separate the energies severely affected by photo-electric
absorption from those little affected (as can been seen in
fig.~\ref{fig:2050xspec}).  We note, however, that the soft band does include
contributions from physically distinct X-ray emission regions-- the heated white dwarf surface, and the
accretion column-- which should be borne in mind in any interpretations of
the lightcurves.  We see
that the flaring does not repeat in each orbit, and that it is far more significant below 1.6 keV. The over-plotted binned data (stepped line) bring out
the average orbital modulations more clearly.  At the higher energies, there appears to be a broad peak, cut unevenly by the eclipse, whereas at
lower energies, the rise in flux to peak does not occur until phase 0.  As we did for \ta, we undertook X-ray spectral fitting to both the
complete dataset, and two phase-selected intervals, here dubbed ``peak'' and ``trough'', excluding the eclipse phase: these regions are
also indicated in fig.~\ref{fig:2050fds}.

%\begin{deluxetable}{llllcclc}
\begin{deluxetable*}{llllcclc}
%\tablenum{1}
\tablecolumns{8} 
\tablewidth{0pt}
\tablecaption{X-ray Spectral Fits for \tb\label{tab:2050xfits}}
\tablehead{ \colhead{Model} &\colhead{reduced} & \colhead{$N_H$} & \multicolumn{2}{c}{Partial covering}&\colhead{$kT_{\rm BB}$} &\colhead{$kT$(s)}\\
  &$\chi^2$   &\colhead{$\times10^{20}$}& \colhead{$N_H$ } & \colhead{frac.}&\colhead{(eV)}&  \colhead{(keV)} \\
 &\colhead{(d.o.f.)}&cm$^{-2}$&\colhead{($\times10^{22}$cm$^{-2}$)}&&&
}
\startdata
bremss &0.99 (418)&$4.3\pm0.3$&\nodata&\nodata&\nodata&$166^{+33}_{-40}$\\
\mek&1.01 (418)&$4.8\pm0.3$&\nodata&\nodata&\nodata& $80^{+0}_{-5}$\\
2 $T$ \mek&1.00(416)&$7.4\pm0.3$&\nodata&\nodata&\nodata&$(8.08^{+0.1}_{-0})\times10^{-2}$\\
&&&&&&$80^{+0}_{-17}$&\\
BB + \mek&0.99 (416)& $5.7^{+0.5}_{-0.4}$&\nodata&\nodata&$20^{+8}_{-6}$&$80^{+0}_{-9}$\\
BB + bremss.&0.98 (416)& $5.6^{+0.4}_{-0.8}$&\nodata&\nodata&$22^{+1}_{-10}$&$100^{+32}_{-19}$\\
2 $T$ \mek\ + BB& 0.99 (412)& $5.1^{+0.5}_{-0.7}$& $1.4^{+0.8}_{-0.6}$& $0.14^{+0.45}_{-0.05}$&$32^{+1}_{-5}$&$1.3^{+1.3}_{-0.5}$\\
&&&&&&$59^{+21}_{-15}$\\
\cutinhead{{\em Joint fit to phase-selected ``peak'' and ``trough'' intervals:}}
%\cutinhead{test}
2 $T$ \mek\ + BB & 0.88 (417)&  $4.9^{+0.5}_{-0.7}$&none (pk)& none (pk) &$28^{+10}_{-13}$&$1.0^{+0.3}_{-0.2}$\\
&&& $1.3^{+0.4}_{-0.3}$(tr)&$0.35\pm{+0.03}$(tr)&&$41^{+5}_{-13}$\\
\enddata
%\tablenotetext{a}{}
%\tablenotetext{b}{Unabsorbed flux in the 0.01--10 keV range, in units of $10^{-13}$\ergsqcmsec.}
%\tablenotetext{c}{Model parameters without quoted uncertainties have been fixed at the value given.}
\end{deluxetable*}
%\end{deluxetable}

For the complete dataset acceptable fits (see Table~\ref{tab:2050xfits})  were found for two component models with Galactic absorption consistent with the maximum line-of-sight
column, with a optically thin thermal component (hot bremsstrahlung or \mek\ equally) plus a soft blackbody.  However, the temperature of the
hotter component at \til80 keV is higher than typically found in CVs. We then investigated the phase-selected spectra seeking a single model
requiring the fewest varying parameters.  The final model consists of a two-temperature thermal component plus a soft
blackbody, with partial covering absorption. In the light of this success, we also fit the complete dataset with the same model; in all cases we found physically
plausible temperatures
and absorption columns.   The dominant contribution to the observed absorbed
flux (98\% in 0.01--10 keV range) is from a hot \mek\ with kT$\approx40$ keV, but $F-$tests confirm
the importance of the soft blackbody to provide the softest X-ray flux (at 93\% confidence), and the cool \mek\ to account for significant unresolved line emission
at \til 1keV (98\%). The change in the spectrum from peak to trough is in part due to increased partial covering absorption: the peak spectrum
requires no absorption in excess of the Galactic column, while for the trough we find a 35\% covering column with
$N_H=1.3\times10^{22}{\rm cm}^{-2}$; and also to a decrease by 25\% in the normalization of
the hot \mek; the blackbody and cool \mek\ normalizations remain unchanged. The
fully unabsorbed 0.01--10 keV flux amounts to $5\times10^{-12}\ergsqcmsec$, with
 40 percent contributed by the soft blackbody. In fig.~\ref{fig:2050xspec} we show the fits to the phase-selected
spectra, together with a break-down of the three components; for clarity we only
show data from the pn, though we used pn and the two MOS spectra when fitting.

\begin{figure}[!tb]
\resizebox{.47\textwidth}{!}{\rotatebox{0}{\plotone{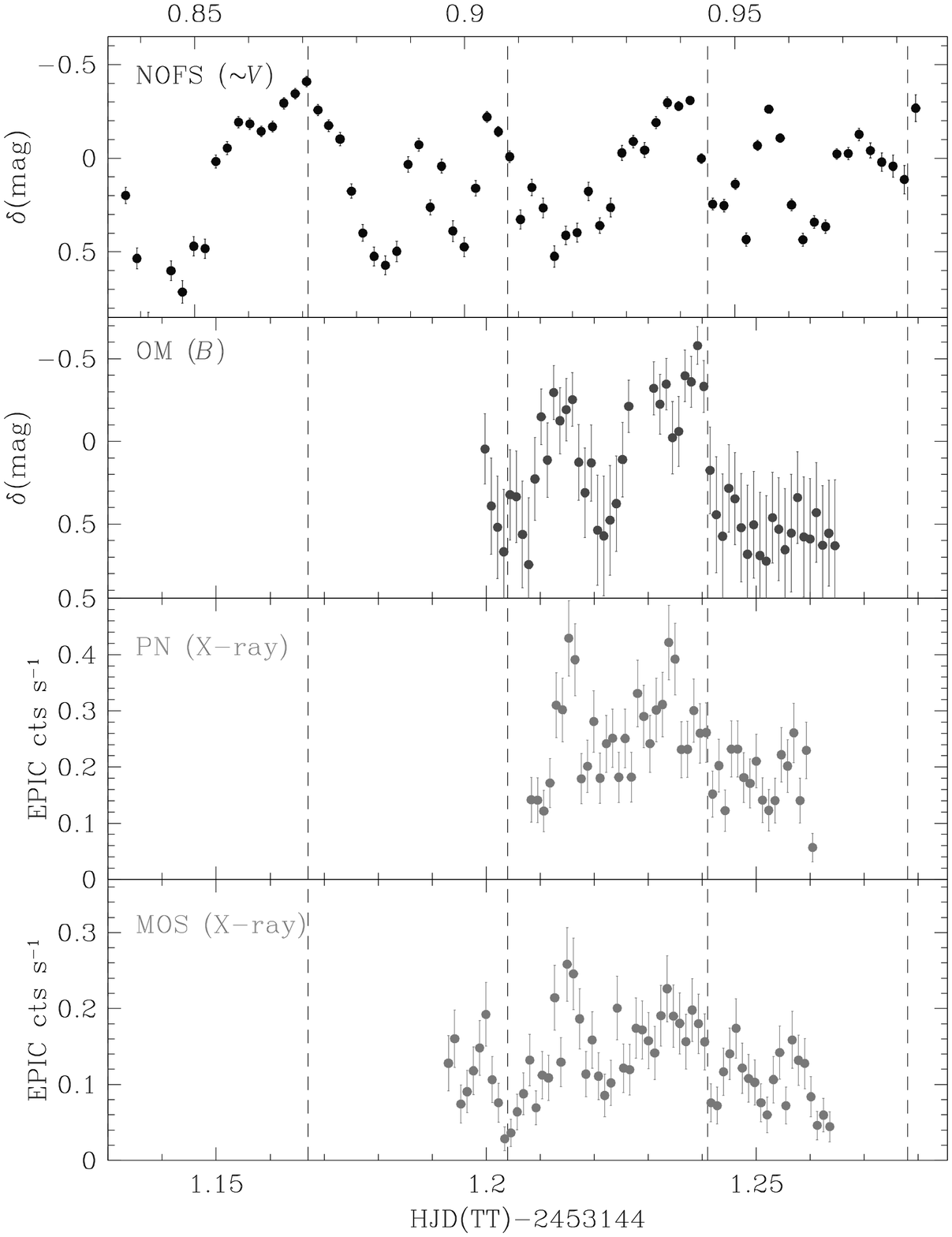}}}
%\resizebox{.47\textwidth}{!}{\rotatebox{0}{\plotone{2101_lcs.eps}}}
\caption{Lightcurves for \tc\ from NOFS (optical) and \sat\ (optical and
  X-ray). The lower three panels share a common time axis, while the earlier
  NOFS observation differs and is marked on the top axis.  The vertical bars
  mark the times of (arbitrary) phase 0.0 and 0.5 for a 107 min period
  (detected in the NOFS lightcurve). \label{fig:2101lcs}}
\end{figure}

\begin{figure}[!tb]
\resizebox{.47\textwidth}{!}{\rotatebox{0}{\plotone{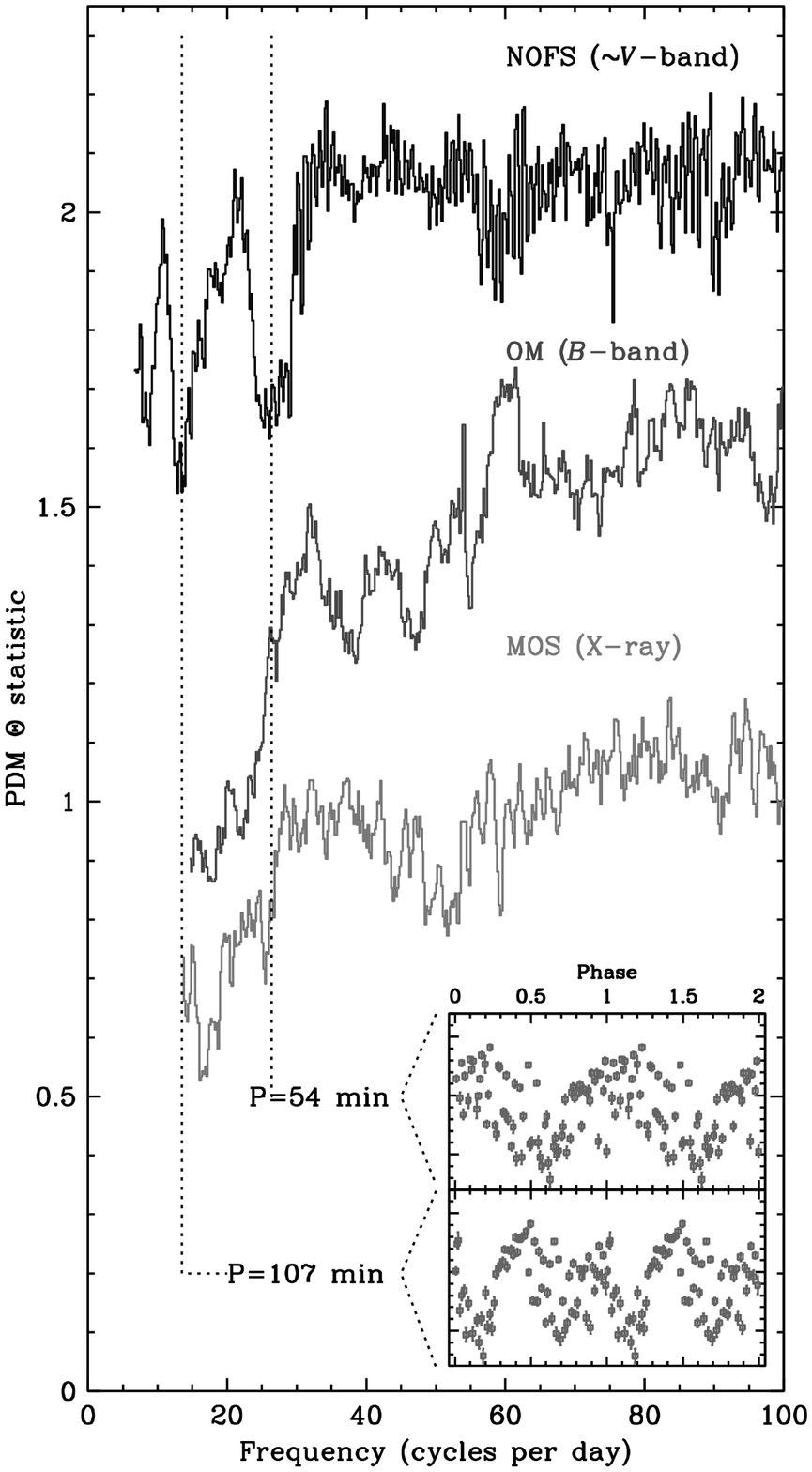}}}
%\resizebox{.47\textwidth}{!}{\rotatebox{0}{\plotone{2101_temp.eps}}}
\caption{Phase Dispersion Minimization periodograms for the MOS X-ray and NOFS and OM optical lightcurves of \tc. The successive curves have
  been offset vertically by 0.5 for clarity, and have been truncated at the low
  frequency end at the inverse of the respective lightcurve time span. The inset shows the NOFS lightcurve
  folded on its candidate periods, 54 and 107
  min. The ordinate is in relative ~$V$ magnitudes, with a 0.5 mag interval between large tick marks. \label{fig:2101temp}}
\end{figure}

\subsection{\tc}
With no previously known periods, either spin or orbital, our first step was to examine
the lightcurves (see fig.~\ref{fig:2101lcs}) for periodicities. Unfortunately, the \sat\ data span only \til1.5 hr, and even the NOFS lightcurve is only 3.5 hrs long, limiting
our sensitivity to any modulations with periods much in excess of 2
hours. Moreover, the variability in both X-ray and optical bands appears
extremely complex. The longest, NOFS
optical, curve appears to show two distinct humps at around 0.87 and 0.94
(truncated HJD(TT)), but of course even here only two cycles are present. Running a phase dispersion minimization 
\citep[PDM,][]{stell78} period-folding search on this light curve, we find three signals, but the lowest frequency one is close to
the inverse of the data time span, and is therefore unreliable.  This leaves
minima at around 13.5 d$^{-1}$ (107 min) and 26.4d$^{-1}$ (55 min), which are
close to being harmonically related to each other.  In figure~\ref{fig:2101temp} we show
the PDM for the NOFS lightcurve and the lightcurve folded on the two candidate
periodicities. The fold on the longer period confirms the repeatability of the
aforementioned humps: only at phases (arbitrary) 0.25--0.5 do the two cycles
agree; but the fold
is also somewhat double-humped, with possibly a second interval of higher
flux at around 0.9, although here there is a large amount of scatter. Folding on
54 min the points line up along an approximately sinusoidal curve, but again
there is significant scatter about this mean at all phases.  Furthermore, for
this shorter period we can usefully interrogate the OM and MOS light curves:
their PDMs are also plotted in fig.~\ref{fig:2101temp}; no strong signals appear
around 54 min.

% [, although the MOS PDM does turn over at \til18$^{-1}$ (\til80 min),
% indicating that if one folds this lightcurve on a period a little shorter than
% its time span (70 min), short segments at the start and finish apparently
% repeat. This is of course a very marginal result, but interestingly, a set of
% time-resolved optical spectra yield a possible period in the radial velocity curve at about 84 min (G\"ansicke, private communication).] Cut??

\begin{deluxetable*}{lllll}
%\tablenum{1}
\tablewidth{0pt}
\tablecaption{X-ray Spectral Fits for \tc \label{tab:2101xfits}}
\tablehead{ \colhead{Model} &\colhead{reduced} & \colhead{$N_H$} & \colhead{$kT_1$ or  } & \colhead{$kT_2$ (keV) }\\
  &\colhead{$\chi^2$ (d.o.f.)}  &\colhead{$\times10^{20}$cm$^{-2}$}&\colhead{$kT_{\rm max}$(keV)}&\colhead{ or $\alpha$}}
\startdata
\mek& 0.82 (95) & 7.1 (fixed)\tablenotemark{a}& $9\pm1$&\nodata\\
\mek& 0.80 (94) & $8.4^{+1.4}_{-0.7}$& $8\pm2$&\nodata\\
2 $T$ \mek& 0.77 (92) &  $8.6^{+1.2}_{-1.1}$&  $9.5^{+3.2}_{-1.6}$&$1.0\pm0.2$\\
multi-$T$ \mek & 0.75 (93)  &$8.6^{+1.4}_{-0.9}$&  $25^{+4}_{-5}$&$1.2^{+0.4}_{-0.2}$\\
\enddata
\tablenotetext{a}{The absorbing column is fixed to that estimated from dust maps, given by the FTOOL {\tt nH} }
\end{deluxetable*}

Returning to fig.~\ref{fig:2101lcs}: in order to consider the NOFS periods further
we have over-plotted vertical bars to mark phase 0.0 and 0.5 of a 107 min period,
arranging the time axes for the \sat\ and NOFS observations to yield phase
agreement (only 4 cycles elapse between the two sets of observations).  We see
that the repeatable NOFS humps agree in phase and approximate form with a hump
in the OM lightcurve. However, at other phases agreement is lost, in particular
another peak precedes the major hump in the OM, possibly
correlated with a similar X-ray peak in fact. Ignoring the high frequency
flickering in the X-ray lightcurves, there is possibly a slower variation on the
\til100 min timescale, though of course we have no way to test whether this
behaviour actually repeats.

\begin{figure}[!tb]
%\resizebox{.47\textwidth}{!}{\rotatebox{-90}{\plotone{2101_xspecfit.eps}}}
\resizebox{.47\textwidth}{!}{\rotatebox{-90}{\plotone{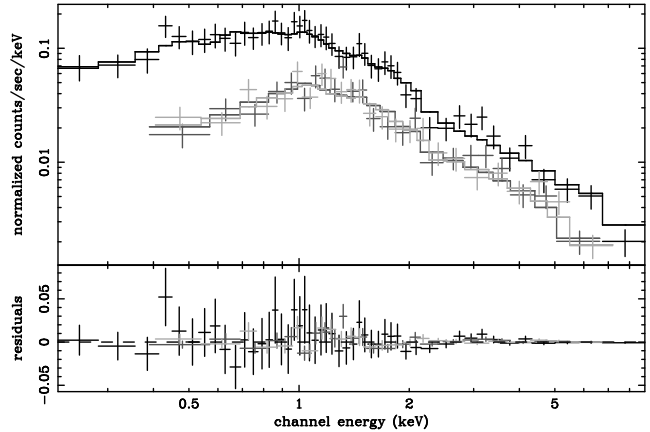}}}
\caption{\sat/EPIC spectra (points) and  best fit model (line) for \tc: {\it black}--pn; {\it dark grey}--MOS1; {\it light grey}--MOS2. This
  model 
  consists of a multi-temperature optically thin plasma model ({\sc cemekl}), absorbed by $N_H=7.5\times10^{19}$ cm$^{-2}$.  In the lower panel the
  residuals to the fit are plotted. \label{fig:2101xspec}}
\end{figure}

In the absence of any decisive periods, we simply fit models to the complete X-ray spectrum.  These are detailed in Table~\ref{tab:2101xfits}.
A range of emission components from  a single fairly cool \mek\ ($kT=9$keV), to a multi-temperature version with $kT_{\rm max}=25$keV provide
adequate fits, all with a single absorbing component with $N_H\approx8\times10^{20}$cm$^{-2}$ consistent with the Galactic line-of-sight maximum. An example
fit to the spectra is shown in fig.~\ref{fig:2101xspec}; the
resulting unabsorbed 0.01--10 keV flux is $1.7\times10^{-12}\ergsqcmsec$.

\section{Classifying the three sources}
\label{sect:disc}
\subsection{Nature of \ta}
All the observational evidence indicates that \ta\ is a short period IP. A
classic signature of IPs is the presence of multiple periodicities in their
lightcurves.  For \ta\ we find that both its optical and X-ray fluxes are strongly modulated at a short 49 min
period, and there is evidence for an additional optical modulation at 4 hrs. This
latter periodicity also appears in the H$\alpha$ radial velocity curve.
However, more extensive photometric and spectroscopic runs are needed to
confirm these, and hence secure the determination of the orbital period. The
identification of the 49 min modulation with the spin period of an asynchronously rotating white
dwarf is much more secure: the characteristic energy dependence of the X-ray
flux modulation, and our simple phase-selected spectral fitting confirm that the
variation in flux is dominated by changes in the local absorbing column. This is
exactly
as seen in many other IPs, as a result of the extended accretion curtain
intersecting our line-of-site to the X-ray emitting accretion column. 
% This simple model can also be invoked to explain the phase-lead of the X-ray
% maximum (or minimum) to that of the optical (specifically the simultaneous $B$
% band from the OM). 

%In the accretion curtain model **ref** the modulation of the opticval flux
%arises from the chnaging a[scets of the optically thick curtainSince we see
%only a single rotational period, and not its beat with the orbit, the reprocessing site for the optical emission must be the disk itself

The underlying X-ray emission can be fit with that emitted by a two-temperature
optically thin plasma, i.e. bremsstrahlung.  No doubt this simple model only
approximates what must be a complex multi-temperature shock region in the accretion
column, however the hotter component with $kT\approx60$keV indicates the upper
limit to the temperatures therein; again this value is typical of IPs.

Assuming that $P_{\rm orb}=4$hr, we find a ratio of $P_{\rm spin}/P_{\rm
  orb}=0.2$, placing \ta\ well-within the group of so-called conventional IPs,
  as defined by $0.25>P_{\rm spin}/P_{\rm orb}>0.01$ and $P_{\rm orb}>4$hr
  \citep{nort03}. In all respects the observed properties of \ta\ are consistent
  with a garden-variety IP.

\subsection{Nature of \tb}
The observational evidence favors a polar classification for \tb.  One spectropolarimetric observation
found significant circular polarization that increases to $v\sim3\%$
for $\lambda > 8000$~\AA.  This result suggests
that the strongest cyclotron harmonics may lie in the near-IR, and
thus that the magnetic field is relatively low, $B \simlt 30$~MG, or
that the specific accretion rate in the cyclotron-emitting portions
of the funnel(s) is relatively low, $\dot m \simlt 1$~g cm$^{-2}$,
or both.  No cyclotron harmonics are evident in either the total flux
or circular polarization spectra, consistent with a high temperature
shock.  Therefore, SDSS J2050$-$05 is probably not a low-accretion
rate polar \citep{schw02,szko03b,schm05b}, where mass-transfer appears to occur not via Roche-lobe
overflow, but rather by efficient magnetic capture of the secondary
star's stellar wind.

In support of the above interpretation, the X-ray spectra of SDSS
J2050$-$05 are well-fit by a two-temperature thermal plasma
model (\mek), but in addition the data
require a soft blackbody component.  In this case, the two \mek\ components have
temperatures of 1 and 41 keV respectively, once again likely indicating the
limits of the 
range of temperatures present in the optically thin emitting plasma, both quite   typical
for polars. The blackbody has a
temperature of about 30 eV, again within the observed range for polars.
In the
``peak'' region, this cool blackbody component, arising from the heated surface
of the white dwarf, contributes about 40 percent of the unabsorbed 0.01-10
keV flux.  In terms of the energy balance, assuming an X-ray albedo for the
white dwarf surface of $a_X=0.3$ and that the cyclotron contribution to the energy losses
is negligible, one finds $L_{\rm BB}/L_{\rm br}\approx\pi f_{\rm
  BB}(1-a_X)d^2/2\pi f_{\rm br}(1+a_X)d^2\approx0.2$, which given uncertainties
in estimating the unabsorbed soft flux is comparable to recent measures of the
energy balance in other
actively accreting polars.

As an eclipsing system, the
orbital period is very well-established.  In the X-ray lightcurves we again see
eclipses, but the remainder of the variability is complex.  It is dominated by
aperiodic flickering behavior, which is strongest in the softer X-ray
flux (below 1.6 keV).  In an attempt to discover any underlying modulation on
the orbital period, we created phase-binned light curves averaging over the three
cycles of data.  What remains appears quite typical of a polar, in which one
accretion pole is always visible (i.e. no white dwarf self-eclipse).  This produces
emission modulated on the
orbitally-synchronized white dwarf spin period, since the projected area of the emitting region changes with phase. The fact that only the normalization of the hotter
thermal emission component appears to vary between ``peak'' and ``trough''
spectra is probably simply owing to poor statistics for the much smaller
contributions from the cooler \mek\ and soft blackbody components. The shape of
the average, non-eclipse
harder X-ray ($E>1.6$keV) lightcurve is indeed approximately sinusoidal, whereas
in the soft band the lightcurve does not rise until after the eclipse, i.e. its flux
remains low in the phase range $\phi\approx0.7-0.95$. The energy-dependence of
this feature suggests that
photo-electric absorption may be the cause.  Indeed, the spectrum for the
$\phi\approx0.4-0.95$ ``trough'' phase interval does require a fairly high
column, which partially covers the X-ray emission. Such pre-eclipse
energy-dependent dipping structure is seen in many other eclipsing polars
e.g. EP Dra, HU Aqr and UZ For \citep{rams04c,schw01,sirk98}. In the high-quality
soft X-ray lightcurves of HU Aqr, there is one broad dip centered on phase 0.7
and a narrower feature at around 0.9.  These are interpreted as arising from
the accretion column obscuring the nearby heated photosphere of the white dwarf,
while the more distant accretion stream accounts for the sharper dip, indeed the
width and azimuth of this dip can be related to the geometry of the region where
the stream threads onto the white dwarf's magnetic field. Unfortunately, given the
quality of our own data (both lacking the high signal-to-noise and afflicted by
the flaring) we cannot identify the dipping components, and hence probe the
accretion flow in greater detail; and of course from our spectral fitting we
know that the soft flux arises both from the white dwarf and various parts of
the accretion column, which would further complicate a more detailed analysis.

\subsection{Nature of \tc}
Although originally identified as a candidate mCV, given the strength of its
high-excitation optical emission lines, the follow-up observations of \tc\ we
have presented do not seem to
confirm this classification.  We find no strong modulation of its X-ray flux,
quite unlike any polar, for instance \tb.  The X-ray spectrum also does
not require any absorption in excess of the estimated Galactic column,
effectively ruling out a typical IP.  The X-ray emission is describable by a
variety of thermal plasma (\mek) models, though with a noticeably lower maximum
temperature ($kT\approx10$ or at most 25 keV) than is found in most mCVs or
indeed in our fits to \ta\
or \tb.  This cooler plasma is in fact far more
typical of disk-accreting systems. Only the SW Sex stars show \ion{He}{2} with
strength comparable to the magnetics; both the lack of absorption in the line
profiles and their barely detectable radial velocities could be accounted for by
a low inclination to our line-of-sight. 

\section{Conclusions}
We have reported on observations allowing classification of three new SDSS CVs
showing prominent 
\ion{He}{2}: \ta, \tb\ and \tc.  Each represents a different class of
CV, known to exhibit such high excitation emission lines. 

\ta\ is a clear example of an IP.  We find two periodicities at about 4 hr and
49 min, arising respectively from the orbit and asynchronously
spinning white dwarf.  The X-ray spectrum comprises a multi-temperature thermal
plasma, with a maximum of 60 keV, modulated by phase-dependent local
absorption. This likely originates from obscuration by the accretion curtains
formed by the accreting material as it flows between the disrupted disc and the magnetic
poles of the white dwarf.

\tb\ represents the most highly magnetized of the three, a
fully spin-orbit synchronized polar, with an orbital period of 1.57 hr. In this
case no disk forms, and all accretion is
channeled along the field lines to a single dominant pole. Owing to its high
inclination, there is a phase interval when the accretion stream obscures
the X-ray emitting region, but most of the modulation on the spin/orbit appears
consistent with simply its changing aspect.  The X-ray emission has two
components: the dominant one is characteristic
of the post-shock bremsstrahlung-cooled optically thin plasma of the accretion column, here with
temperatures ranging from 1 keV to 40 keV; this also heats the surface of the white
dwarf leading to additional soft blackbody (30 eV) emission. 

In \tc, we suggest that the combination of high accretion rate
and lower magnetic field strength allows
disk-mediated accretion, but also the formation of \ion{He}{2} emission lines in the
optical spectra, i.e. if it were at higher inclination it would appear as a
classic SW Sex star.  Its X-ray spectra indicate a rather cool thermal plasma
(\til10 keV) more typical of disk systems, and an unobscured line of sight,
inconsistent with an IP. From only weak modulation (unlike a polar) of its X-ray and optical lightcurves
we suggest an orbital period in the 100 min range.  

\acknowledgments
The authors wish to thank the anonymous referee for their thorough review and
useful comments, that helped improve the paper.  This work was supported by \sat\ grant NNGO4GG66G to the University of
Washington and is based on observations obtained with \sat, an ESA science mission with instruments and contributions directly funded by ESA
Member States and the USA (NASA).  G. Schmidt acknowledges the support of 
NSF grant AST 03-06080 and P. Szkody AST 02-05875.

 Funding for the SDSS and SDSS-II has been provided by the Alfred P. Sloan
 Foundation, the Participating Institutions, the National Science Foundation,
 the U.S. Department of Energy, the National Aeronautics and Space
 Administration, the Japanese Monbukagakusho, the Max Planck Society, and the
 Higher Education Funding Council for England. The SDSS Web Site is
 http://www.sdss.org/.

The SDSS is managed by the Astrophysical Research Consortium for the Participating Institutions. The Participating Institutions are the American Museum of Natural History, Astrophysical Institute Potsdam, University of Basel, Cambridge University, Case Western Reserve University, University of Chicago, Drexel University, Fermilab, the Institute for Advanced Study, the Japan Participation Group, Johns Hopkins University, the Joint Institute for Nuclear Astrophysics, the Kavli Institute for Particle Astrophysics and Cosmology, the Korean Scientist Group, the Chinese Academy of Sciences (LAMOST), Los Alamos National Laboratory, the Max-Planck-Institute for Astronomy (MPIA), the Max-Planck-Institute for Astrophysics (MPA), New Mexico State University, Ohio State University, University of Pittsburgh, University of Portsmouth, Princeton University, the United States Naval Observatory, and the University of Washington. 

%\thebibliography

%\bibliographystyle{/Users/homer/home/papers/bst/apj}
%\bibliography{/Users/homer/home/papers/bib/aas}

\end{document}